\documentclass{aa} % 
\usepackage{graphicx}
\usepackage{txfonts}
\usepackage{psfrag,pstricks,epsfig,tabularx}
\usepackage{epstopdf}
\usepackage{ulem}
\usepackage{gensymb}

\begin{document} 

\title{Disparity among low first ionization potential elements}
\titlerunning{Disparity among low FIP elements}
\author{Verena Heidrich-Meisner \inst{1}, Lars Berger \inst{1}, and Robert F. Wimmer-Schweingruber \inst{1}}

\authorrunning{V. Heidrich-Meisner, L. Berger, and R. Wimmer-Schweingruber}
%\authorrunning{V. H.-M., T. P., M. K., L. B.,  R. W.-S.}

\institute{Christian Albrechts University at Kiel, Germany,
  \email{heidrich@physik.uni-kiel.de}
}
%todo: add address!

\date{}

\abstract{%context
  The elemental composition of the solar wind differs from the solar photospheric composition. Elements with low first ionization potential (FIP) appear enhanced compared to O in the solar wind relative to the respective photospheric abundances. This so-called FIP effect is different in the slow solar wind and  the coronal hole wind. However,  under the same plasma conditions, for elements with similar FIPs such as Mg, Si, and Fe, comparable enhancements are expected.
}{%Aims
  We scrutinize the assumption that the FIP effect is always similar for different low FIP elements, namely Mg, Si, and Fe.
}{%methods
  Here we investigate the dependency of the FIP effect of low FIP elements on the O$^{7+}$/O$^{6+}$ charge state ratio depending on time, that is the solar activity cycle, and solar wind type. 
  In addition, we order the observed FIP ratios with respect to the O$^{7+}$/O$^{6+}$ charge state ratio into bins and analyze separately  the respective distributions of the FIP ratio of Mg, Si, and Fe for each  O$^{7+}$/O$^{6+}$ charge state ratio bin. 
}{%results
  We observe that  the FIP effect shows the same qualitative yearly behavior for Mg and Si, while Fe shows significant differences during the solar activity maximum and its declining phase.  In each year, the FIP effect for Mg and Si always increases with increasing O$^{7+}$/O$^{6+}$ charge state ratio, but for high O$^{7+}$/O$^{6+}$ charge state ratios the FIP effect for Fe shows a qualitatively different behavior.  During the years 2001--2006, instead of increasing with the O$^{7+}$/O$^{6+}$ charge state ratio, the Fe FIP ratio exhibits a broad peak {or plateau}. In addition, the FIP distribution per O$^{7+}$/O$^{6+}$ charge state bin is significantly broader for Fe than for Mg and Si. 
}{%conclusions
  {These observations support the conclusion that the elemental fractionation is only partly determined by FIP. In particular, the qualitative difference behavior with increasing O$^{7+}$/O$^{6+}$ charge state ratio between Fe on the one hand and Mg and Si on the other hand is not yet well explained {by models} of fractionation.}
}


   \maketitle
%
%________________________________________________________________

   \section{Introduction}

The solar elemental composition has been under investigation for several decades \citep[]{feldman1992elemental,schmelz-etal-2012, meyer1985solar}. 
The photospheric elemental composition is typically assumed to be constant with the solar activity cycle and is used as the reference for the elemental composition observed in active regions \citep[]{baker2015fip}, in coronal loops \citep[]{del2003solar}, and in the solar wind \citep[]{feldman1992elemental,schmelz-etal-2012, meyer1985solar}. It is well known that some elements are depleted or enriched in the solar wind compared to the photosphere. Since elements  with similar first ionization potentials (FIPs) show a similar elemental composition in the solar wind relative to the photosphere, this is called the ``FIP effect'' and the underlying mechanism behind this effect separates ions from neutrals, and is thus assumed to act preferentially on ions rather than on neutral atoms. It is still unknown whether low FIP elements are enhanced in the solar wind compared to the photosphere or whether instead high FIP elements are depleted. Both interpretations are consistent with the observations.

{Several theoretical approaches have been explored in modeling the FIP 
effect, including quasi-thermal processes of inefficient Coulomb 
drag/gravitational settling/diffusion \citep[]{vsteiger-and-geiss-1989,marsch1995element,bochsler2000solar,pucci2010elemental,peter1996velocity,peter1998element}, heating 
by coronal ion cyclotron waves \citep[]{schwadron1999astrophysical}, chromospheric 
reconnection \citep[]{arge1998modelling}, electric currents \citep[]{feldman2003elemental}, and most recently and most promisingly, the ponderomotive force arising as Alfv{\'e}n waves propagate through or reflect from the chromosphere \citep[]{doschek2015anomalous,laming2012non,laming2015fip,dahlburg2016ponderomotive,laming2017first}.}

The FIP effect is more pronounced in the slow solar wind than in the fast coronal hole wind \citep[]{feldman1992elemental,schmelz-etal-2012, meyer1985solar}. 
Since they have similar FIPs, Mg (with a FIP of $7.646$ eV), Si (with a FIP of $8.151$ eV), and Fe (with a FIP of $7.870$ eV; values taken from \citealt{benenson2006handbook}) are typically assumed to express the same FIP effect  in the slow solar wind and in the coronal hole wind. This is  consistent with long-term averages of the respective elemental abundances \citep[]{schmelz-etal-2012} and the dominant models of the FIP effect \citep[]{laming2015fip}. {However, as discussed in detail in \citet{laming2017determining}, the FIP effect is not the only cause of fractionation in the solar wind. }  {\citet{pilleri2015variations} have already observed considerable fractionation between {low} FIP elements.}

{As visible in Figure 1 of \citet{schmelz-etal-2012} and discussed in detail in \citet{reames2018fip}, a}nother interesting fractionation case is S (with a FIP of 10.36 eV), which fractionates as a high FIP element in the closed loop solar corona, but is observed with a higher FIP ratio in the solar wind.

One of the main science goals of the upcoming Solar Orbiter mission \citep[]{muller2013solar} is to identify the source regions of the slow solar wind. In tune with the general scheme of Solar Orbiter to heavily exploit combined observations of its instruments, a coordinated observation of the elemental abundance by the SPectral Imaging of the Coronal Environment (SPICE) and the Heavy Ion Sensor (HIS) is intended as the main tool for investigating potential slow  {solar} wind source regions. Both instruments can determine the strength of the FIP effect for several elements, and coordinated observations of the same plasma packages are planned. However, this relies on the assumption that in particular the low FIP elements Mg, Si, and Fe show highly correlated behavior even on short timescales. {Here, we utilize the available observations of the  {S}olar {W}ind {I}on {C}omposition Spectrometer (SWICS) on the {Advanced Composition Explorer (ACE)}} to scrutinize this assumption and investigate the differences in the behavior of these low FIP elements. 

%\verena{look for examples that use a lowFIP value}

\begin{figure*}\centering
  \includegraphics[height=.9\textheight]{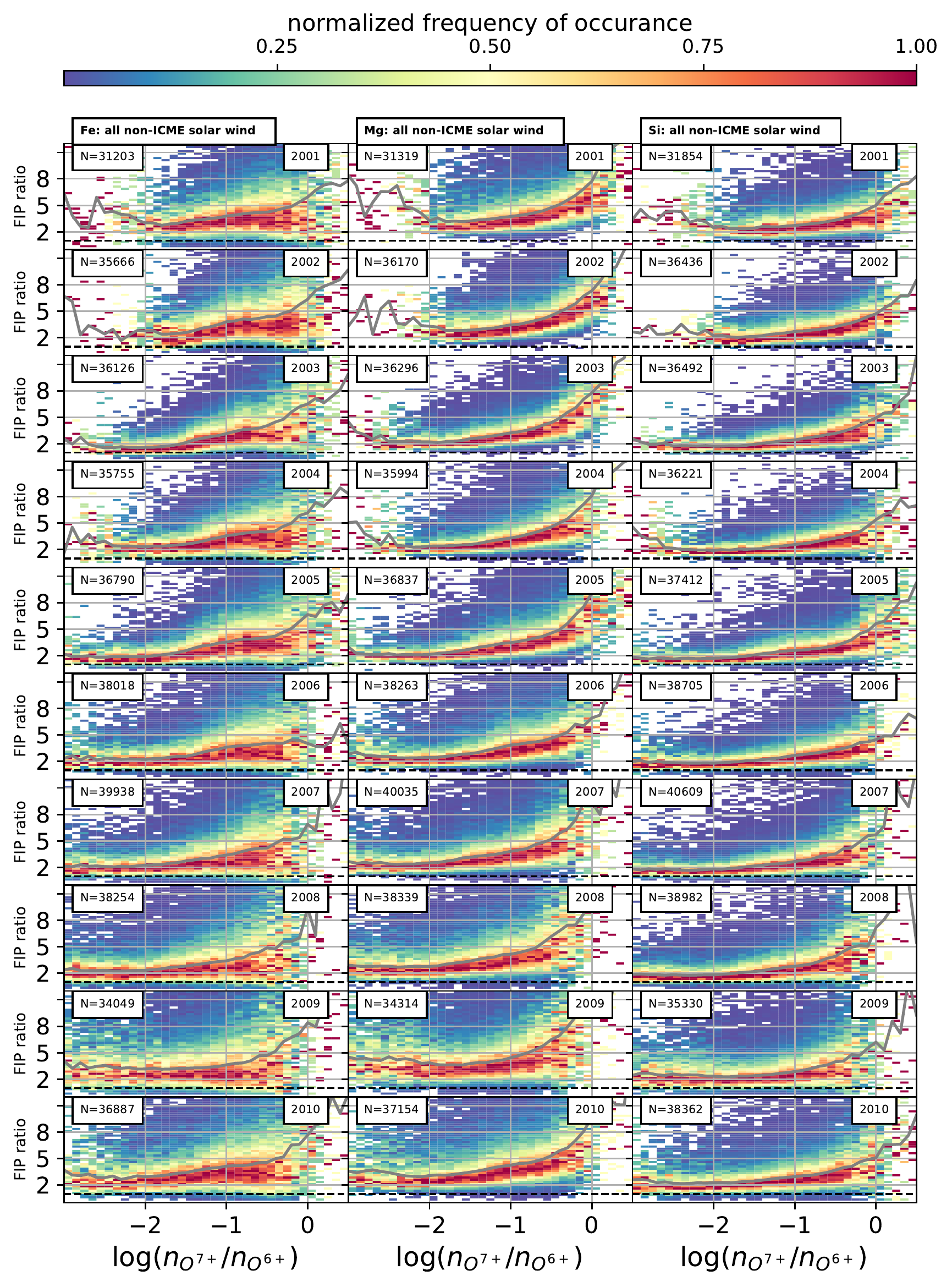} 
  \caption{\label{fig:allfip} FIP ratios per year from 2001 to 2010. The columns show the FIP ratios for Fe, Mg, and Si. Each row corresponds to the same year. In each panel{,} a 2D histogram of all 12-minute observations of FIP ratios vs the $\log\left(\frac{n^{O7+}}{n^{O6+}}\right)$ from the respective year are shown. Each vertical slice is normalized to its maximum.  The number of contained data points is given in the inset
of  each panel. The dashed horizontal line indicates a FIP ratio of one. {The gray solid line marks the median of the FIP ratio in each respective O charge state ratio bin.}}
\end{figure*}

\begin{figure*}
  \includegraphics[width=\textwidth]{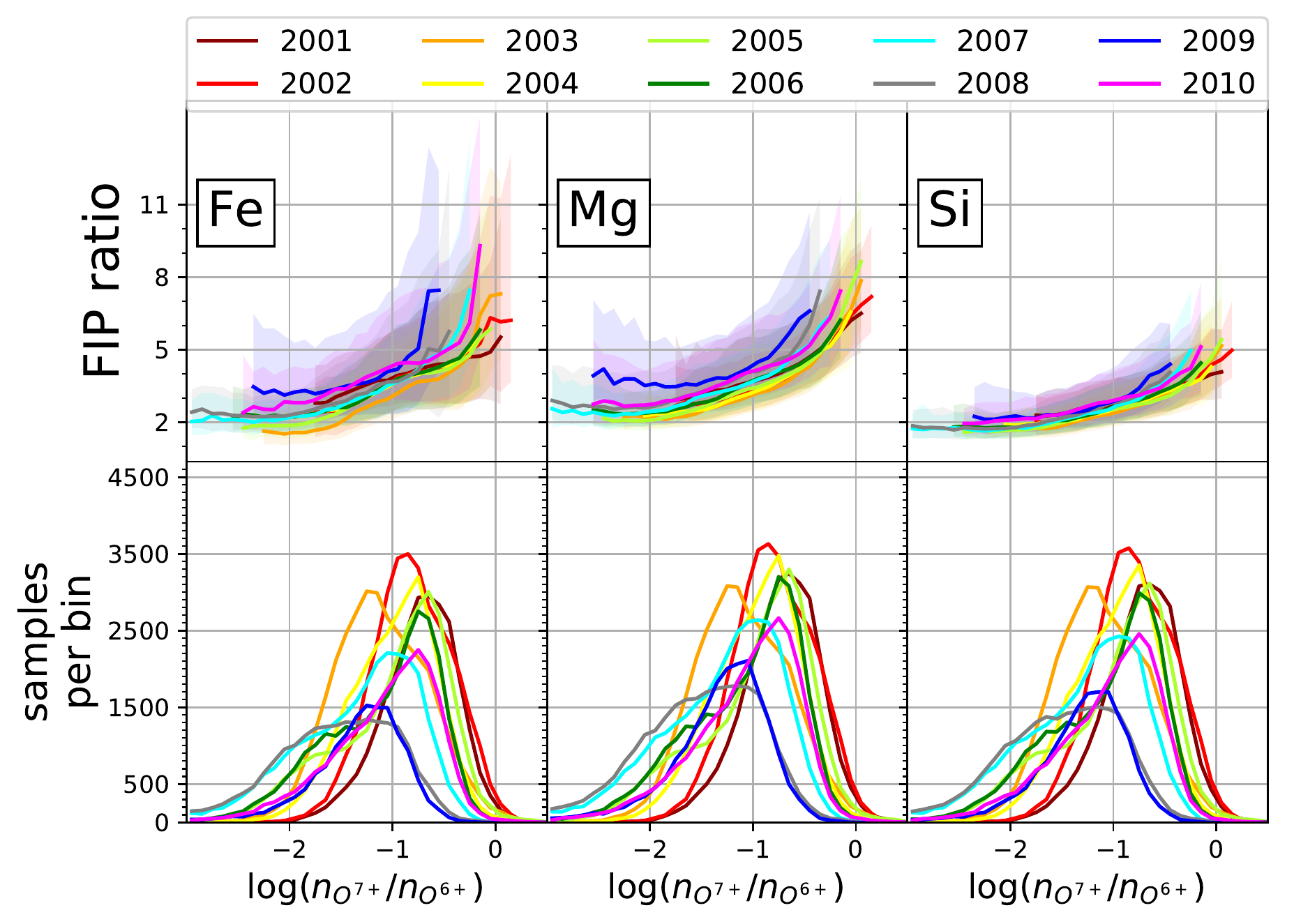}
  \caption{\label{fig:medianfip} Median {FIP} ratios per year from 2001 to 2010 for Fe, Mg, and Si. The median is given only if at least 100 samples were observed in the respective bin. {The colored shading indicates the 15.9th and 81.1th percentile for each year.} In the lower panel{s} the number of samples per O charge state bin and year are shown for Fe, Mg, and Si.}
\end{figure*}

\begin{figure*}\centering
  \includegraphics[height=.9\textheight]{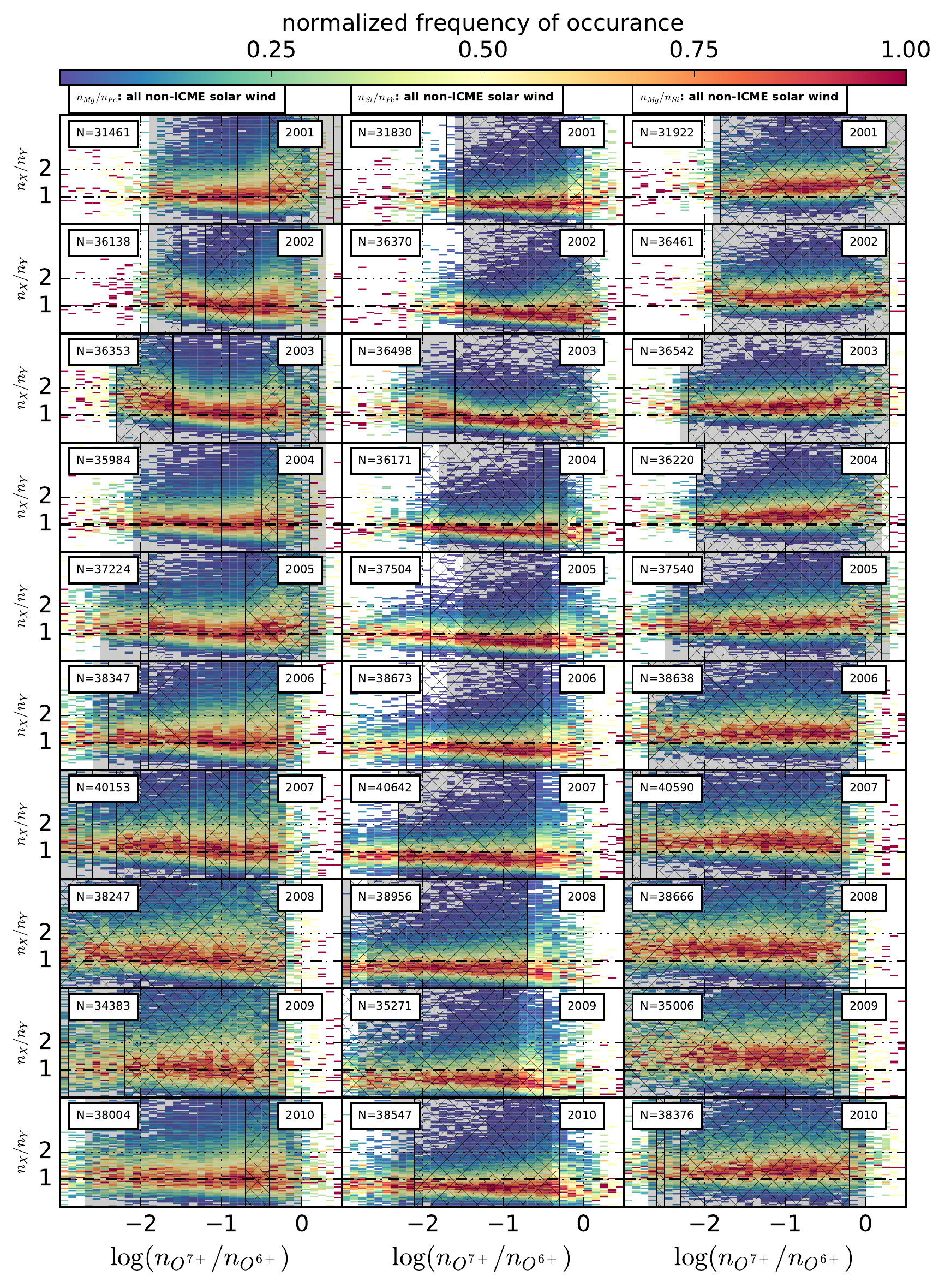}
  \caption{\label{fig:allRatios} Relative densities for pairs of low FIP elements. The columns correspond to the density ratios $Mg/Fe$, $Si/Fe$, and $Mg/Si$. As in Figure~\ref{fig:allfip}, each panel shows a 2D histogram of 12-minute observations for one year normalized for each {O} charge state ratio bin. The insets again specify the number of data points contained in each panel. Gray shading indicates bins with significantly different mean densities and the  cross-hatching marks bins where the standard deviation of the two distributions are significantly different. {The dashed horizontal line marks an abundance ratio of one.}}
\end{figure*}

\begin{figure*}
  \includegraphics[width=\textwidth]{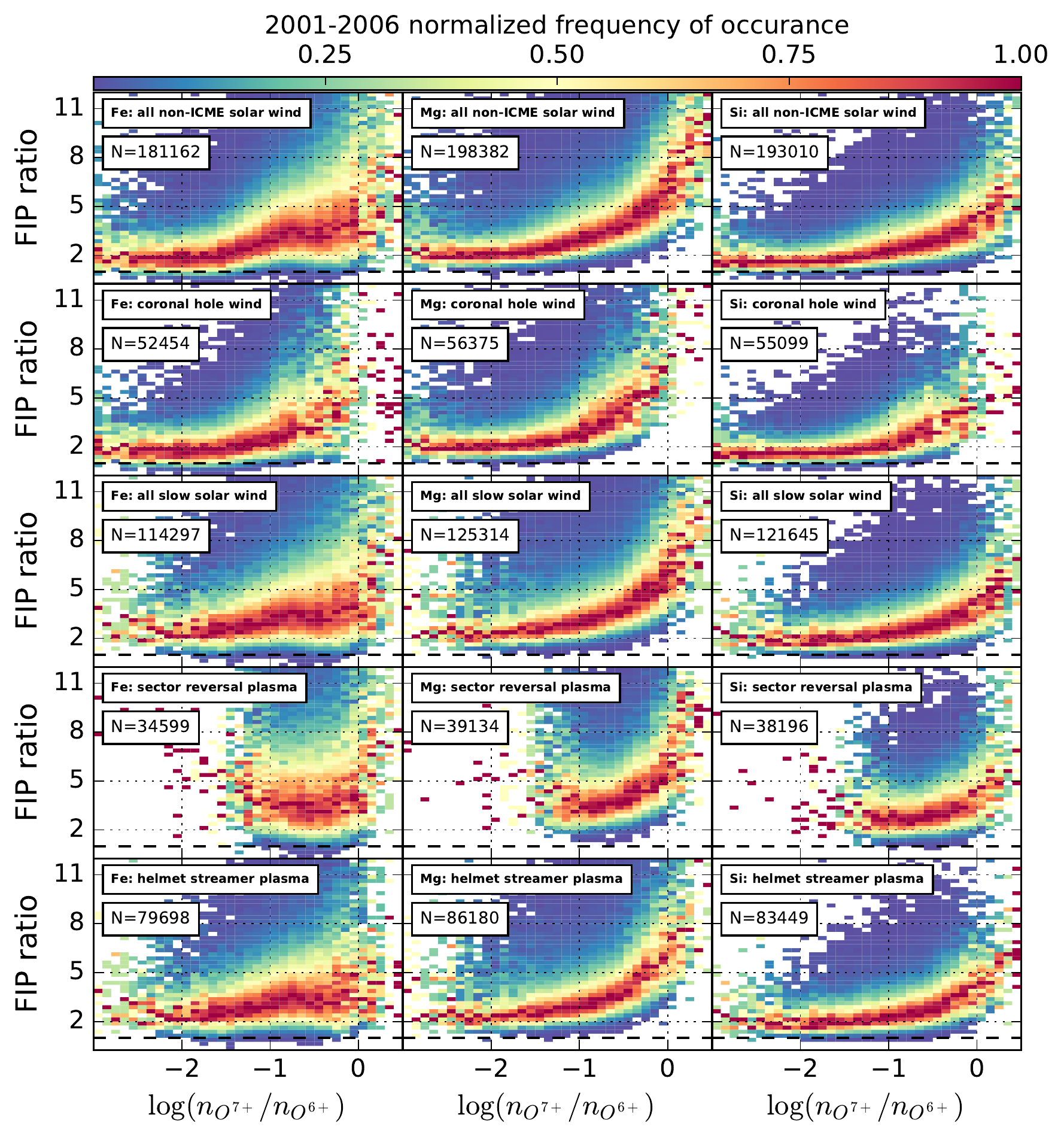}
  \caption{\label{fig:swfipmax}FIP ratios per solar wind type in a similar format to that in Figure~\ref{fig:allfip} accumulated for the years 2001--2006. Each row shows a different solar wind type (based on \citealt{xu2014new}), from top to bottom: all non-ICME solar wind, coronal hole wind, all slow solar wind, sector reversal plasma, and helmet streamer plasma.}
\end{figure*}

\begin{figure*}
  \includegraphics[width=\textwidth]{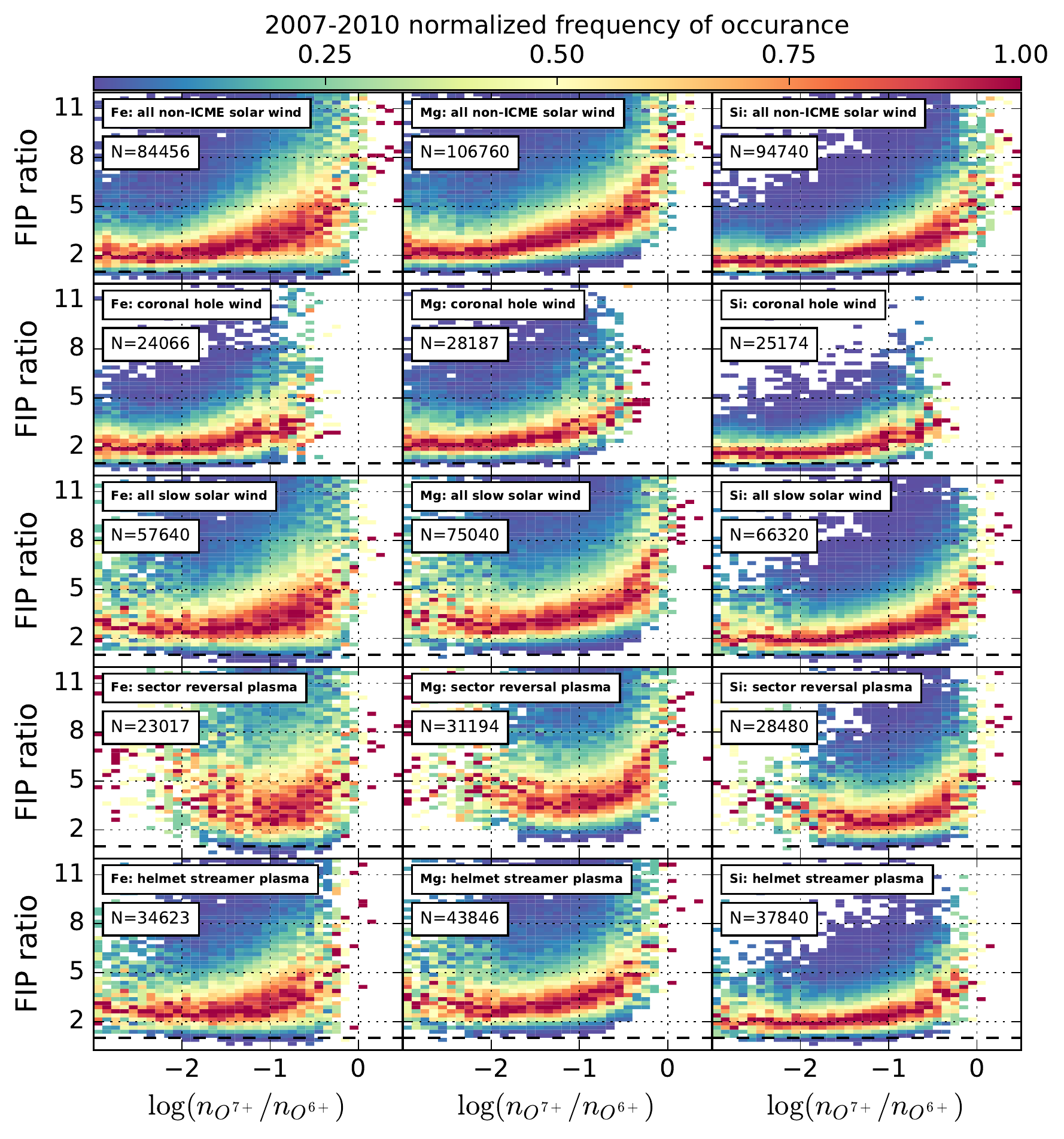}
  \caption{\label{fig:swfipmin}FIP ratios per solar wind type in the same format as in Figure~\ref{fig:swfipmax} accumulated for the years 2007--2010.}
\end{figure*}

\section{Data selection}
We are interested in the elemental composition of low FIP elements in the solar wind and apply our analysis to ten years of data from the {S}olar {W}ind {I}on {C}omposition spectrometer {(SWICS, \citealt{gloeckler-etal-1998})} on the {Advanced Composition Explorer (ACE)} . The solar wind proton plasma parameters are taken from the Solar Wind Electron, Proton, and Alpha Monitor (ACE/SWEPAM) \citep[]{mccomas1998solar} and magnetic field observations from the magnetometer  ACE/MAG \citep[]{smith1998ace}.

The ionic composition is derived from the SWICS Pulse Height Analysis (PHA) words as described in \citet{berger2008velocity}. We use the native 12-minute time resolution of ACE/SWICS which results in at most 43800 data points in non-leap years and 43920 in leap years.

The elemental abundance of O, Mg, Si, and Fe is taken as the sum of the respective most prominent charge states, namely O$^{6+}$, O$^{7+}$, Mg$^{7+}$, Mg$^{8+}$, Mg$^{9+}$, Mg$^{10+}$, {Mg$^{11+}$}, {Mg$^{12+}$}, Si$^{7+}$, Si$^{8+}$, Si$^{9+}$, Si$^{10+}$, Si$^{11+}$, {Si$^{12+}$}, Fe$^{7+}$, Fe$^{8+}$, Fe$^{9+}$, Fe$^{10+}$, Fe$^{11+}$, Fe$^{12+}$, Fe$^{13+}$, {Fe$^{14+}$}, {Fe$^{15+}$}, and  {Fe$^{16+}$}.  Furthermore, to reduce statistical noise we require for each element a minimum of ten counts distributed over all respective ions. All data points that violate this condition are disregarded. However, since this induces a bias against very thin solar wind conditions, we also verified that the qualitative effect discussed in this paper remains if data points with fewer counts per element are included as well. 
We refer to the ratio of the abundance of an element $X$ with density $n_X$ relative to O in the solar wind $\left(\left(\frac{n_{X}}{n_{O}}\right)_{sw}\right)$ divided by the respective photospheric ratio $\left(\left(\frac{n_{X}}{n_{O}}\right)_{photo}\right)$  taken from \citet{grevesse1998standard} as the $X$ FIP ratio $f(X)=\frac{ \left(\frac{n_{X}}{n_{O}}\right)_{sw}}{\left(\frac{n_{X}}{n_{O}}\right)_{photo}}$. %\verena{todo: use newer photospheric values?}

To characterize the solar wind type, we employ the four-type solar wind categorization scheme from \citet{xu2014new}. This heuristic scheme distinguishes between coronal hole wind, two types of slow solar wind, sector-reversal plasma and helmet streamer plasma, and ejecta plasma. This solar wind categorization scheme does not rely on the charge state composition, but only on (proton) plasma properties of the solar wind. To exclude interplanetary coronal mass ejections (ICMEs) we use {two ICME lists, the \citet{jian2006properties,jian2011comparing}  list and the \citet{richardson2010near}  list, instead of the ejecta category. We exclude all time intervals from the data set that are considered as belonging to an ICME in either of the lists. Nevertheless, unidentified ICMEs are probably still included in the data set.} We refer to the union of the two slow solar wind types as ``all slow'' solar wind. 
We repeated our analysis based on the composition based solar wind categorization from \citet{zhao2010comparison}, and observed the same qualitative behavior.

\section{{Fractionation} of low FIP elements}
For Ne, a dependence of the elemental abundance on the solar cycle has been observed \citep[]{von2015solar}. Under the assumption that Ne does not behave in a fundamentally different way from  other elements in the Sun and the solar wind, this motivates us to independently analyze the FIP effect for Mg, Si, and Fe for each year.

Figure~\ref{fig:allfip} provides yearly overviews of the observed FIP ratios of Mg, Si, and Fe from 2001 to 2010 to the {logarithm of the} respectively observed O$^7+$/O$^{6+}$ charge state ratio (we refer to this as the O charge state ratio only in the following). In the context of the Genesis mission, \citet{pilleri2015variations} carried out a similar study and observed a decrease in the Fe and Mg FIP ratio with increasing solar wind speeds. Since the O charge state ratio is likely to be better suited to sort by solar wind type than the solar wind speed \citep[]{zhao2010comparison}, we instead use the O charge state ratio as reference.

Although all three elements show in all years a wide spread in the observed FIP ratios, the most frequently occurring FIP ratios range between $1$ and $4$. For Mg and Si, each year shows a similar pattern: with increasing O charge state ratio, the FIP ratio increases as well. This is consistent with the observations that the slow solar wind shows a stronger FIP effect than the coronal hole wind (as summarized in \citealt[]{schmelz-etal-2012} {and \citealt{pilleri2015variations}} with 2--4 for low FIP elements in slow solar wind and 1-2 for coronal hole wind). For Fe, however, the trend is different{, in particular,} during the solar activity maximum (for example in 2004) than during the solar activity minimum (for example in 2009). In 2009, the Fe FIP ratio shows no trend (or only a weak one)  with increasing O charge state ratio. In 2004, the Fe {FIP} ratio exhibits a broad peak around $\log\left(\frac{n_{O^{7+}}}{n_{O^{7+}}}\right)\sim-0.7$. Also, the Fe FIP ratio distribution in most O charge state bins {appears} broader than for Mg and Si. 
When using the solar wind speed as a reference value, as in \cite{pilleri2015variations}, both the Mg and Fe FIP ratio continue to decrease with increasing solar wind speed.

Figure~\ref{fig:allfip} further illustrates that the {O} charge state ratio changes over the solar cycle. During the solar activity  maximum, higher values of the O charge state {ratio} are reached, while during solar minimum, lower O charge state {ratio} values are observed. Thus, the part {of the regime} of the O charge state range which exhibits a broad peak in the Fe FIP ratio from 2001 to 2006, is only infrequently observed in the following years. Thus, from this figure it cannot be excluded that the Fe FIP effect always shows the same qualitative behavior in all years, but this is not visible in 2007--2010 because of insufficient statistics for this very slow solar wind. Although most ICMEs are filtered out from the data set, any ICME list can be expected to be incomplete. Since ICMEs are typically also associated with high charge states and are more frequent during the solar activity maximum, undetected ICMEs {probably} contribute to the broad peak in the Fe FIP ratio.  Also, as discussed in, for example, \citet{klecker2009observation}, the density of high Fe charge states (with charges > 13)  is significantly enhanced during ICMEs. However, since during normal solar wind conditions their contribution to the elemental density can be neglected, {these high Fe charge states mainly contribute to the Fe elemental abundance for very high O charge state ratios, that is $\log\left(\frac{n^{O7+}}{n^{O6+}}\right)>-0.5$. Thus, for corresponding slow solar wind with $\log\left(\frac{n^{O7+}}{n^{O6+}}\right)>-0.5$, Figure~\ref{fig:allfip} is probably not representative.} It is interesting to note that typically ICMEs are also associated with very high FIP ratios, whereas here the Fe FIP ratio is lower than expected for high O charge state ratios. Furthermore, for Mg and Si, the same high O charge state ratios are shown in 2001--2006 as for Fe, but the FIP ratios of Mg and Si nevertheless continues to increase with increasing O charge state ratio. {For each O charge state {ratio} bin, the distribution of FIP ratios of Mg, Si, and Fe are asymmetric. Therefore, the median per O charge state {ratio} bin, which is indicated with solid gray lines in Figure~\ref{fig:allfip}, is not a good proxy for the most frequently observated FIP ratios, but instead always overestimate them.}

In the upper panels in Figure~\ref{fig:medianfip}{,} the yearly median of the FIP ratio for Mg, Si, and Fe is shown. In the lower panels,  the number of samples per bin are given for each
year. For all three low FIP elements, the median FIP ratio is lower in the years 2001--2006 and higher in the years 2007--2010. This indicates a solar cycle dependence of the FIP effect for all three elements. However, this effect is small compared to the variability of the observations as indicated by the percentiles in Figure~\ref{fig:medianfip}. {Unlike the most frequently observed Fe FIP ratios, the median of the Fe FIP ratios behaves more like  the respective yearly medians of the Mg and Si FIP ratios. Nevertheless, the slope of the median Fe FIP ratio  for high O charge ratios is smaller during the solar activity maximum than during the solar activity minimum. This is not the case for the year 2003,  probably because of the unusually high fraction of coronal hole wind during this year (see, e.g., Figure~2 in \citealt{pilleri2015variations}).} 

Figure~\ref{fig:allRatios} compares the FIP ratios of each pair of low FIP elements (Mg versus Fe, Si versus Fe, and Mg versus Si) for each year. With the help of statistical tests,  for each O charge state {ratio} bin we investigated whether the distributions of FIP ratios of each pair of low FIP elements are likely to follow the same or different distributions. The shaded/hatched areas mark significant differences between the two respective mean values (gray shading, $p<0.05$ according to the Kolmogorov--Smirnoff test; \citealt[]{kolmogorov1933sulla,smirnov1948table}) and empirical standard deviations (cross-hatching, $p<0.05$ based on Levene's test; \citealt[]{olkin1960contributions}). Both statistical tests are suitable for non-normal distributions and were only applied if a minimum of 100 samples were observed per O charge state {ratio} bin. Both the mean and the standard deviation of Mg and Si are significantly different for all years and over almost the complete range of O charge state ratios. The distributions of Si and Fe are significantly different as well for intermediate O charge state ratio in all years, but show more similarities both for low and high O charge state ratios.  For Fe and Mg, the mean of the respective distribution is also significantly different for most O charge state {ratio} bins, but for {most} years the empirical standard deviations of the respective distributions are comparable for intermediate and low O charge state ratios (which correspond  mainly to the coronal hole wind). {Except for 2002}, for high O charge state ratios (which correspond  mainly to the slow solar wind), and thus for the regime that exhibits the unexpected behavior in the Fe FIP ratio, both mean and empirical standard deviation of Fe and Mg differ significantly. It is also notable that the bulk of the ratio distributions is not at one (except for {high O charge state {ratio} bins in} the cases of Mg and Fe). {In any O charge state {ratio} bin, t}here is on average less Si observed than Fe, and {for most O charge state {ratio} bins} more Mg than Fe. This is consistent with the observations in \citet{reames2018fip} who observed higher Mg than Fe abundances for co-rotating interaction regions, the inter stream solar wind, and the coronal hole wind. However, \citet[]{pilleri2015variations} instead found higher abundances for Fe than for Mg. We suspect that the difference arises from a respective omission or inclusion of Ca ions in the fitting process. Including Ca probably leads to some counts that were caused by Fe ions to be incorrectly assigned to Ca which leads to an underestimated Fe abundance. Excluding Ca from the fitting process correspondingly probably leads to incorrectly assigning counts caused by Ca ions to Fe which leads to an overestimation of Fe. One or both effects can result in the different observed relative abundances. However, in all cases these differences of the mean FIP ratio are within the respective error bars. %

Although the distribution of the Mg and Si FIP ratios in Figure~\ref{fig:allfip} appear similar, Figure~\ref{fig:allRatios} shows that for most O charge state {ratio} bins, the respective distributions of Mg and Si have significantly different means and standard deviations.

To ensure that the effect discussed here is not simply caused by different mixtures of different solar wind types in each year, in the following we sort the observations by solar wind type.
Figure~\ref{fig:swfipmax} and Figure~\ref{fig:swfipmin} show only solar wind from 2001--2006 and 2007--2010, respectively, but the observations are also separated by solar wind type according to the \citet{xu2014new} scheme. During 2001--2006 for the coronal hole wind, all three elements show the same qualitative increase in the FIP ratio with increasing O charge state ratio; however,  the FIP effect is most pronounced for Mg and weakest for Si. 
Even for the coronal hole wind, the Fe FIP ratio distribution {appears} broader than for Mg and Si during this time period. For the slow solar wind, in particular for helmet streamer plasma, Fe shows a qualitatively different behavior from that of  Mg and Si with increasing O charge state ratio during 2001--2006. {For Fe, s}ector reversal plasma exhibits a particularly broad FIP {ratio} distribution and no clear dependence of the FIP effect on the O charge state ratio. During 2007--2010, the distribution of the observed Fe FIP ratios is also broader than for Mg and Si, but for each solar wind type the qualitative behavior is more similar. 

\section{Discussion and conclusion}
The observations reveal that the long-term behavior of the elemental composition is qualitatively different for Fe and for Mg and Si. While for Mg and Si the FIP ratio always increases with increasing O charge state ratio in all considered years and for all solar wind types, this is not the case for Fe. For Fe, the FIP ratio reaches a {plateau or} broad peak {for logarithms of the} O charge state ratio between 0.1 and 1 during the years 2001--2006. As also supported by Figure~\ref{fig:swfipmax} and Figure~\ref{fig:swfipmin}, the differences are most pronounced for slow solar wind. That the Fe FIP ratio behaves differently from Mg and Si mainly in the slow solar wind during the declining phase of the solar activity maximum can indicate that a {different fractionation or release mechanism} is dominant during this period than during the solar activity minimum. %

In addition, the Fe FIP distribution per O charge state {ratio} bin is {typically} broader than the respective distributions for Mg and Si. Based on the available models of the FIP effect, such a qualitative difference between the FIP ratios of low FIP elements is not expected. Thus, we believe that a full explanation of the observations discussed here requires refinement and improvement of the available models of the FIP effect. 

The observations presented here clearly indicate that the assumption that all low FIP elements show a highly correlated FIP effect is not always valid, and thus that {one or more} other fractionation mechanisms play an important role.

\begin{acknowledgements}
      Part of this work was supported by the Deut\-sche
      For\-schungs\-ge\-mein\-schaft (DFG)\/ project number
      Wi-2139/11-1\enspace. {We thank the referee,
        Martin Laming, for his insightful, helpful, and detailed remarks. We also thank the science teams of ACE/SWEPAM, ACE/MAG, as well as ACE/SWICS for providing the respective level 2 and level 1 data products.}
\end{acknowledgements}
   \bibliographystyle{aa} % style aa.bst
   \bibliography{aa} % your references Yourfile.bib
%   \clearpage

\end{document}